\documentclass[12pt,preprint]{aastex}
\def\xmm{{\sl XMM}-Newton}
\def\psra{PSR~J0437--4715}
\def\psrb{PSR~J2124--3358}
\def\psrc{PSR~J1024--0719}
\def\psrd{PSR~J0034--0534}
\def\fp{f_{\rm p}}
\def\ed{\dot{E}}
\def\lnon{L^{\rm nonth}}
\def\nh{n_{\rm H,20}}
\def\tbb{T^\infty_{\rm bb}}
\def\rbb{R^\infty_{\rm bb}}
\def\tpc{T_{\rm pc}}
\def\rpc{R_{\rm pc}}
\def\tpcc{T_{\rm pc}^{\rm core}}
\def\rpcc{R_{\rm pc}^{\rm core}}
\def\tpcr{T_{\rm pc}^{\rm rim}}
\def\rpcr{R_{\rm pc}^{\rm rim}}

\def\lbol{L_{\rm bol}}
\def\lbolpc{L_{\rm bol}^{\rm pc}}
\def\epc{\eta^{\rm pc}}

\def\lnon{L^{\rm nonth}}

\newcommand{\gapr}{\raisebox{-.6ex}{\mbox{
$\stackrel{>}{\mbox{\scriptsize$\sim$}}\:$}}}
\newcommand{\lapr}{\raisebox{-.6ex}{\mbox{
$\stackrel{<}{\mbox{\scriptsize$\sim$}}\:$}}}

\begin{document}

\shortauthors{V.\ E. Zavlin}
\shorttitle{X-rays from four millisecond pulsars}
\title{\xmm\ observations of four millisecond pulsars}
\author{Vyacheslav E.\ Zavlin\altaffilmark{1}\altaffiltext{1}{
Space Science Laboratory, NASA MSFC SD50, Huntsville, AL 35805;
vyacheslav.zavlin@msfc.nasa.gov
}}
\begin{abstract}
I present an analysis of the
\xmm\ observations of four millisecond pulsars,
J0437--4715, J2124--3358, J1024--0719, and J0034--0534.
The new data provide strong evidence of thermal
emission in the X-ray flux detected from
the first three objects. This thermal component is best interpreted
as radiation from pulsar polar caps covered with a nonmagnetic
hydrogen atmosphere.
A nonthermal power-law component, dominating
at energies $E\gapr 3$ keV, can also be present in the detected X-ray
emission. For \psra, the timing analysis
reveals that the shape and pulsed fraction of the pulsar
light curves are energy dependent. This, together with 
the results obtained from the phase-resolved spectroscopy,
supports the two-component (thermal plus
nonthermal) interpretation of the pulsar's X-ray radiation.
Highly significant pulsations 
have been found in the X-ray flux of PSRs J2124--3358 and J1024--0719.
For \psrd, a possible X-ray counterpart of the radio pulsar
has been suggested. The inferred properties of the detected
thermal emission are compared with predictions of radio pulsar models.

\end{abstract}
\section{Introduction}
Of more than 1500 currently known rotation-powered (radio) 
pulsars\footnote{
According to the catalog provided by the Australia Telescope
National Facility (ATNF),
{\tt http://www.atnf.csiro.au/research/pulsar}, for the middle of 2005.},
122 possess very short spin periods $P<10$ ms.
These millisecond pulsars 
are generally believed to be very old neutron
stars (NSs), spun up by accretion in binary systems,
with characteristic ages $\tau=P/2\dot{P}\sim 0.1$--10 Gyr
and low surface magnetic fields 
$B_{\rm surf}\propto (P\dot{P})^{1/2}\sim 10^8$--$10^{10}$ G 
(e.g., Taylor et al.\ 1993). 
As millisecond pulsars are intrinsically faint at optical 
wavelengths and most of them
reside in binary systems with optically brighter
white dwarf companions, X-ray energy range is the main source of
information on these objects outside the radio band.
So far firm X-ray detections have been reported
for nine millisecond (non-variable) pulsars, either solitary or in
binaries without interaction with companions. They are given in Table~1 
(this list does not include X-ray emitting millisecond pulsars 
in the global clusters 47~Tuc and NGC~6397 because of a lack of
a detailed information on their X-ray properties ---
see, e.g., Grindlay et al.\ 2002).

X-ray emission from radio pulsars consists of two 
different
components, thermal and nonthermal (see, e.g., Pavlov et al.\ 2002,
and Kaspi et al.\ 2006 for reviews). 
The nonthermal component is 
described by a power-law (PL) spectral model and attributed to
radiation emitted from pulsar magnetosphere, whereas the thermal
emission can originate from either the whole surface of a cooling NS or 
small hot spots around the magnetic poles (polar caps; PCs)
on the star surface, or both.
As predicted by virtually all pulsar models, these PCs can be heated
up to X-ray temperatures ($\sim 1$ MK) by relativistic
particles generated in pulsar acceleration zones.
In the case of millisecond pulsars, 
the entire surface at a NS age of $\sim 1$ Gyr is too cold, 
$\lapr 0.1$ MK, to
be detectable in X-rays (although it may be seen in $UV/FUV$
--- see \S\,2). Therefore, only nonthermal and/or
thermal PC radiation is expected to be observed in X-rays
from these objects.
The pulsars presented in Table~1 can be divided in two 
distinct groups
(see also Kuiper \& Hermsen 2004). 
The first group consists of 
PSRs B1937+21, B1957+20, J0218+4232, and
B1821--24, with high rotational energy losses, 
$\ed>10^{35}$ erg s$^{-1}$, and estimated X-ray luminosities,
$L_X\gapr 10^{32}$ erg s$^{-1}$.
Their X-ray emission is of a nonthermal 
origin, with pure PL spectra of photon index $\Gamma=1.5$--2
and pulsed profiles with strong narrow peaks and large pulsed 
fractions, $\fp\gapr 50\%$ (see references in Table~1). 
Probably, PSR~J0751+1807 can also be associated with this group 
because
its spectrum is better fitted with a PL model of $\Gamma\approx 1.6$.
Three other pulsars, J0437--4715, J0030+0451
and J2124--3358, with lower spin-down energies, 
$\ed<10^{34}$ erg s$^{-1}$, belong to the second group.
X-ray emission from these NSs
shows significant contribution of a 
thermal component, with smoother pulsations and lower pulsed fractions.
The properties of X-rays detected from PSR~J1012+5307 
are rather uncertain. However, there is an
indication of a thermal component in its X-ray flux.
Current theoretical models are not elaborate
enough to predict in which millisecond pulsars the thermal PC component
would prevail over the nonthermal one. 
It, however, seems certain 
that both the thermal and nonthermal luminosities should 
increase with rotational energy loss $\ed$.
Therefore, only analysis of X-ray 
observations can help in elucidating the radiative properties 
of millisecond pulsars.

The high sensitivity and spectral resolution of $Chandra$ and
\xmm\ provide new opportunities for studying X-ray emission
from faint celestial objects.
This paper presents an analysis and interpretation of
new \xmm\ observations of four millisecond pulsars,
J0437--4715, J2124-3358, J1024--0719, and J0034-0534
(\S\S\,2--5). The obtained results are summarized
and discussed in \S\,6.

\section{\psra}
This is the nearest and X-ray brightest millisecond pulsar known.
It is in a 5.7 d binary orbit with a low-mass white dwarf companion.
Its pulsed X-ray emission was first detected with $ROSAT$ 
in the 0.1--2 keV range (Becker \& Tr\"umper 1993). 
Those data could be fitted with either a single PL
model of $\Gamma\approx 2.5$, suggesting a nonthermal origin
of the detected emission, or a thermal PC model consisting of
two opposite PCs covered with nonmagnetic hydrogen atmosphere\footnote{ 
See Zavlin \& Pavlov (2002) for a recent review on NS atmosphere models.}
and temperature
decreasing from $\sim 1.5$ MK in the PC "core" of about $\sim 0.3$ km
radius to $\sim 0.4$ MK in the PC ``rim'' of a $\sim 3$ km radius
(Zavlin \& Pavlov 1998). Becker \& Tr\"umper (1993) found that
pulsed fraction $\fp$ 
of the pulsar flux increases with photon energy $E$
from about 30\% at $E\sim 0.2$ keV up to about 50\% at 1 keV
(although Becker \& Tr\"umper 1999 found no energy dependence of $\fp$
from a later $ROSAT$ observation of \psra).
On the other hand, thermal emission from NS atmospheres 
is essentially anisotropic, even at low $B_{\rm surf}$, 
with radiation intensities beamed stronger at higher $E$
(Zavlin et al.\ 1996), that can result in an increase of $\fp$
with photon energy.
Therefore, if the energy dependence of $\fp$ found by Becker
\& Tr\"umper (1993) is real, 
it could be regarded as an additional indication 
of the thermal origin of the detected radiation.
The pulsar was also observed with $Chandra$.
Analysis of the combined $ROSAT$ and $Chandra$ data (Zavlin et al.\ 2002)
showed that the 0.1--7 keV spectrum of \psra\ cannot
be described with a simple one-component model and requires two
components: a nonthermal PL spectrum of $\Gamma\approx 2.0$ and 
a thermal PC model similar to that suggested by Zavlin \& Pavlov (1998)
from the $ROSAT$ data alone. The thermal (bolometric) and nonthermal 
(in 0.2--10 keV)
luminosities were
found to be $\lbol\approx 2.3\times 10^{30}$ erg s$^{-1}$
and $\lnon\approx 0.6\times 10^{30}$ erg s$^{-1}$ 
(for a distance $d=0.14$ kpc --- see Table~1),
respectively. In this interpretation, the nonthermal component
dominates at energies $E\gapr 3$ keV. The $Chandra$ data provided
also an energy-integrated pulse profile
with $\fp\approx 40\%$ and the X-ray peak at the
same phase as the radio peak. In addition, \psra\ is
the only millisecond pulsar detected near the optical,
at the wavelengths $\lambda=1155$--1702 (Kargaltsev et al.\ 2004),
that provided an estimate on the temperature
of the entire NS surface, $T_{\rm surf}\approx 0.1$ MK ---
surprisingly high for such an old NS, implying special heating
mechanism(s) to operate in this pulsar.

\subsection{\xmm\ observation}
\psra\ was observed with \xmm\ on 2002 October 9 (orbit 519)
for 68.3 and 67.2 ks effective exposures with the 
EPIC\footnote{European Photon Imaging Camera}-MOS and
EPIC-pn instruments, respectively.
EPIC-MOS1 and MOS2 were operated in Full Frame mode
providing an image of an area of a $\sim 14'$ radius
and a time resolution of 2.6 s. EPIC-pn was in Timing mode.
In this mode only one EPIC-pn CCD
is in use and the EPIC-pn image is
collapsed in one-dimensional (1-D) count distribution in the direction
perpendicular to the CCD read-out, that provides 
a 0.03 ms time resolution suitable
for detecting pulsations in X-ray flux of millisecond pulsars.
Thin filters were used for each EPIC instrument. The EPIC data 
were processed with the most recent 
SAS package\footnote{{\tt http://xmm.vilspa.esa.es}} (ver. 6.1.0).

Figure~1 presents a $20'\times 20'$ image constructed from 
combined EPIC-MOS1 and MOS2 data. 
\psra\ is the second brightest source in the filed of
view. The EPIC-pn field of view is also shown in Figure~1, 
as well as the direction of the CCD read-out. 
It is evident that the satellite position
angle ($138.7^\circ$) chosen for observing \psra\ was not optimal, 
which led to a partial contamination
of the pulsar data in the 1-D EPIC-pn image by the much brighter
neighboring AGN. This resulted in an asymmetric shape of
the 1-D distribution of EPIC-pn counts shown
in the upper panel of Figure~2.
To evaluate the contamination, we constructed three 1-D images 
from the EPIC-MOS data confined within the EPIC-pn field of view 
with the same column width of $4\farcs2$ as in the EPIC-pn CCD: 
from the areas above and below
the short-dashed line in Figure~1, which contain negligible fractions 
of pulsar and AGN quanta (respectively), 
and from the total area. Examination of these 
three distributions plotted  
in the lower panel of Figure~2
shows that the contamination of the pulsar is insignificant
in the CCD columns below \#\,40,
which were used for the analysis below.

\subsection{Spectral analysis}
Both EPIC-MOS spectra of \psra\ were extracted from circles
of a $30''$ radius
centered at the pulsar position, each containing about 82\% of
all detected source counts. Background was estimated from
a few source-free regions.  The extracted spectra 
were binned with at least 30 source counts per bin.
The estimated total count rates of the pulsar are $161\pm2$ and
$167\pm 2$ counts ks$^{-1}$ in the 0.3-10 keV range
for the EPIC-MOS1 and MOS2 detectors, respectively 
(energies below 0.3 keV
were not used because of uncertainties in the EPIC-MOS calibration).
The EPIC-pn spectrum was extracted from the CCD columns \#\,28--39, what
provides about 85\% of all detected source counts. Background
was estimated from the columns \#\,12--23.
The EPIC-pn spectrum was binned with at least 100 
source counts in each bin. The total source count rate was measured to be
$674\pm 5$ counts ks$^{-1}$ in the 0.3--10 keV range (there is no
spectral information below 0.3 keV in the EPIC-pn detector in
Timing mode). All instrument response matrices and effective
areas were generated with the SAS tools {\it rmfgen} and {\it arfgen} 
(these tools were also used in analysis of all \xmm\ data 
discussed in this paper).
The source count rates in the Reflection Grating Spectrometers,
$\sim 20$ counts ks$^{-1}$, are too low for a meaningful analysis
of this data.

Applying a simple one-component model to the three combined EPIC spectra 
resulted in unreasonable model parameters and unacceptable fit quality.
A best single (absorbed) PL fit results in a photon index of 
$\Gamma\approx 3.5$ and absorbing hydrogen column density of
$\nh=n_{\rm H}/(10^{20}\,{\rm cm}^2)\approx 11.6$ (vs. the estimate
$\nh=0.1$--0.3 obtained from independent measurements --- see
Zavlin \& Pavlov 1998), with the minimum value of $\chi^2_\nu=2.47$
(for $\nu=426$ degrees of freedom). Even worse fit is yielded by 
a one-temperature thermal model, e.g., $\chi^2_\nu=7.14$
for a single blackbody (BB) spectrum, because of a large
data excess at energies $E\gapr 2$ keV.  A two-component model,
BB+PL, results in $\chi^2_\nu=1.67$ ($\nu=424$). With account for
problems in  the cross-calibration of the EPIC instruments\footnote{
See Kargaltsev et al.\ (2005) for results of the spectral
analysis of \xmm\ data on the NS RX~J1856.5--3754, as well as
other examples at
{\tt http://xmm.vilspa.esa.es/docs/documents/CAL-TN-0018-2-4.pdf}\,.}
 it can be considered as a reasonably good fit. 
In this model the PL component with $\Gamma=2.9$
provides about 70\% of the observed X-ray flux
($f_X\approx 1.1\times 10^{-12}$ erg s$^{-1}$ cm$^{-2}$
in 0.2--10 keV). However, the obtained
value of $\nh=1.4$ still significantly exceeds the independent estimates
on the hydrogen column density.
In addition, the inferred photon index is much larger than
$\Gamma=1.4-2$ found in nonthermal emission from other radio 
(including millisecond ones) pulsars.
A similar problem arises when one-temperature PC model with
hydrogen atmosphere is applied. Hence, we regard this model
as rather infeasible. The next option 
is a PC model with a nonuniform temperature
distribution combined with a PL component.
Applying the ``core''+``rim'' model for two PCs covered with a hydrogen
atmosphere (Zavlin et al.\ 1996)
and assuming standard NS mass $M=1.4 M_\odot$ and
radius $R=10$ km yields\footnote{
Errors in fitting parameters are given at a 1$\sigma$ level
for one interesting parameter.} 
$\tpcc=1.4\pm 0.2$ MK and $\tpcr=0.52\pm 0.16$ MK (unredshifted
values), $\rpcc=0.35\pm 0.18$ km and $\rpcr=2.6\pm 0.4$ km, 
and $\Gamma=2.0\pm 0.4$,  for $\nh$ varying between 0.1 and 0.3,
with the minimal value of $\chi^2_\nu=1.57$ ($\nu=422$).
Formally, according to an $F$-test, the probability that the spectral 
data on \psra\ require a two-temperature PC model
(instead of a one-temperature thermal component) is 99.99992\%.
This model implies equal angles $\zeta$ 
(between the line-of-sight and pulsar spin axis) and $\alpha$
(between the magnetic and spin axes) of $45^\circ$.
Figure~3 presents a best fit with this two-component model.
The estimated bolometric luminosity of two PCs is 
$\lbol=(3.4\pm0.7)\times 10^{30}$ erg s$^{-1}$,
and the PL luminosity in 0.2--10 keV is
$\lnon=(0.5\pm0.2)\times 10^{30}$ erg s$^{-1}$.
Changing the pulsar geometry (angles
$\zeta$ and $\alpha$) 
as well as the NS mass-to-radius ratio in a plausible
range $M/R=(1.0$--$1.8)\,M_\odot/(10\,{\rm km})$ responsible for the
gravitational effects (redshift and bending of photon trajectories)
increases the errors in the PC parameters, but insignificantly it
affects the inferred thermal luminosity (see Zavlin \& Pavlov
2004 for more details). Generally, the spectral results obtained from the
EPIC data are in good agreement with those derived from 
the previous observations.

Another hypothesis is that the pulsar spectrum is of a pure
nonthermal origin with the PL slope changing somewhere in the
X-ray range. To verify this, a broken PL model was applied to the
EPIC spectra, that yielded a best fit with $\nh=2.3\pm 0.5$,
and photon indices $\Gamma_1=2.4\pm 0.2$ and $\Gamma_2=3.6\pm 0.1$
below and above the break energy $E_{\rm br}=1.05\pm 0.05$ keV
($\chi^2_\nu=1.73$ for $\nu=424$). However, this model significantly 
overpredicts the $UV/FUV$ fluxes detected from \psra\ 
(Kargaltsev et al.\ 2004). This, together with
the large value of the interstellar absorption inferred in 
the fit, makes the broken-PL interpretation rather implausible.

\subsection{Timing analysis}
For the timing analysis, we used 38393 counts  extracted
from the EPIC-pn CCD columns \#\,35--39
in the 0.3-6 keV range chosen to maximize the signal-to-noise ratio
($S/N=163$).
Of those counts, 88\% belong to the pulsar.
The photon arrival times were transformed to the solar system
barycenter with the {\it barycen} tool of the SAS package
(this procedure was also applied to all other timing data
discussed in this paper).
As timing parameters of millisecond pulsars are known to be very stable,
we invoked the binary ephemeris parameters from
Zavlin et al.\ (2002). These parameters used in the $Z^2_n$ test
($n$ is number of harmonics --- see Buccheri et al.\ 1983)
yielded the most significant value $Z^2_n=2170.2$ with $n=2$.
The maximum of $Z^2_{2,{\rm max}}=2199.1$ 
is reached at a frequency $f$ which differs from the radio value 
by $\delta{f}=0.8$ $\mu$Hz,  corresponding to a relative accuracy, 
$\delta{f}/f\simeq 5\times 10^{-9}$,
consistent with that estimated for the EPIC-pn timing\footnote{
See {\tt http://xmm.vilspa.esa.es/docs/documents}}.

We extracted the pulsed profiles of \psra\ in five
energy ranges, 0.3--0.5, 0.5--0.8, 0.8-2, 0.3--2 keV, and
2-6 keV, which are given in Figure~4 together with estimated
values of pulsed fraction $\fp$. The statistical $\chi^2$ test
described in Zavlin \& Pavlov (1998) shows that 
the shapes of the three lower-energy light curves are actually the
same (the probability that they are different is less than 90\%).
However, the 2--6 keV light curve seems to be more symmetric, 
with a peak apparently narrower than those 
in the lower-energy pulse profiles. 
One may also speculate that there is a phase shift of 
$\delta\phi\sim 0.1$ between the peaks at energies below and 
above 2 keV. This apparent difference is 
supported by the same statistical test yielding 
a fairly high probability of 99.99997\% (or a 4.2$\sigma$ significance)
that the shapes of the 0.3--2 and 2--6 keV 
pulse profiles are different.
Another feature in the extracted light curves is the 
change of the source intrinsic pulsed fraction
with energy: $\fp$ increases by about 10\% from 
$\sim 32\%$ at the lowest energies to $\sim 42\%$ at $E\sim 1$ keV,
similar to the energy behavior of $\fp$ 
first found by Becker \& Tr\"umper (1993) from the $ROSAT$ data.
The estimated pulsed fraction at $E>2$ keV is less certain
because of a scantier statistics and 
strong background contamination at higher energies, although
it may be as high as $\sim 65\%$. Generally, 
the energy dependence of the pulsar light curves 
can be attributed to different emission mechanisms producing
the X-ray flux of \psra, as suggested by the thermal-plus-nonthermal 
model.

Figure~5 shows the pulse profiles of \psra\ obtained from 
the $ROSAT$ Position Sensitive Proportional Counter, 
$Chandra$ High Resolution Camera (see Zavlin et al.\ 2002)
and \xmm\ EPIC-pn data (in 0.3--6 keV).
The pulsed fractions in the latter two light curves are 
similar to each other and somewhat larger than that found
in the $ROSAT$ data.
This may be explained by both the properties of the pulsar emission
(increase of $\fp$ with energy)
and the greater sensitivity of the \xmm\ and $Chandra$ instruments 
to higher-energy photons.
The shapes of these three light curves are clearly asymmetric,
with a longer rise. This asymmetry could be caused by contribution
of the nonthermal component (more than 10\% at all energies)
whose peak is shifted in phase with respect
to the pulse of the thermal emission.

As the error in the EPIC-pn absolute timing can be as large
as 0.5 ms, the phasing between the radio and X-ray peaks
was not attempted (this is also one of the reasons why the
zero phases in Fig.~5 are arbitrary).

\subsection{Phase-resolved spectroscopy}
In principle, the EPIC-pn data on \psra\ allow one to perform
a phase-resolved spectroscopy of the pulsar emission.
To do this, we extracted the pulsar spectra  
in five equal intervals between the rotational phases 0 and 1,
with the zero-phase as in Figures~4 and 5, from
the same CCD columns chosen
for the phase-integrated spectrum (\S\,2.2).
The spectra were binned to collect at least 50 source counts per bin.
A simplest and broadly used approach is to fit the phase-resolved
spectra with the same model to determine the dependences of model
parameters on rotational phase. Based on the results of \S\,2.2,
we applied the two-temperature PC (``core''+``rim'')
plus PL model with temperatures of the thermal components
and photon index fixed at the best values derived for
the phase-integrated spectrum. The hydrogen column density
was also fixed at $\nh=0.2$. Figure~6 shows phase dependences
of the three fitting parameters, sizes of the PCs and nonthermal
flux, normalized by the corresponding values found in the
phase-integrated spectral analysis. 
The former two represent the PC areas projected 
onto a plane perpendicular to the line-of-sight of a distant observer. 
As it could be expected, these two dependences 
are very similar to each other and resemble the shape of 
the lower-energy light curves (at $E<2$ keV) shown in Figure~4.
On the other hand, the phase dependence of the nonthermal flux
clearly differs from those of the PC areas, with a 
narrower peak shifted with respect to
the maxima in the PC phase dependences, indicating that
the shift between the peaks in the pulsed profiles of \psra\ 
extracted at energies below and above 2 keV (Fig.~4) is real.
Also, the modulation of the nonthermal flux is rather large,
by about 60\%, pointing to a high intrinsic pulsed fraction
of the nonthermal component.

More detailed information could be obtained from combining the
phase-resolved spectroscopy with modeling the pulsar light curves.
However, such an analysis is beyond the scope of this paper. 

\section{\psrb}
X-ray emission from this solitary pulsar was first detected
with $ROSAT$ (Becker \& Tr\"umper 1999). These data
provided a 4$\sigma$ detection of pulsations of the pulsar
X-ray flux with a pulsed fraction $\fp\sim 33\%$.
\psrb\ was also observed with $ASCA$ (Sakurai et al.\ 2001).
No significant pulsations were found
in those data of a scanty statistics.
The pulsar spectrum could be equally well fitted with either a single
PL spectrum of photon index $\Gamma\sim 2.8$ or a thermal BB 
model of a temperature\footnote{
The superscript ``$\infty$'' stays for redshifted values
as measured by a distant observer.}
$\tbb\sim 3.6$ MK
emitted from an area of an apparent radius $\rbb\sim 0.02$ km
(for a distance $d=0.27$ kpc).
Based on the inferred model parameters, 
Sakurai et al. (2001) concluded that the thermal interpretation
is more plausible.

\subsection{\xmm\ observation}
\xmm\ observed \psrb\ on 2002 April 14--15 (orbit 430) for
68.9 and 66.8 ks effective exposure for the EPIC-MOS and EPIC-pn
detectors operated with medium and thin filters, respectively,
in the same observational modes as for \psra.
Figure~7 shows a $20'\times20'$ combined EPIC-MOS1 and MOS2
image with the pulsar located close to the image center.
With the satellite position angle of $76.7^\circ$ in
this observation, the pulsar emission detected in the EPIC-pn data 
was contaminated by weaker field sources
(see Fig.~7 for the EPIC-pn field of view).
Analysis of the 1-D distribution of EPIC-pn photons shows that the
contamination is substantial
only at energies $E\gapr 2$ keV. Hence, the 0.3--2 keV range
was used for the analysis of the pulsar radiation detected with 
the EPIC-pn instrument.

\subsection{Spectral analysis}
Two EPIC-MOS spectra \psrb\ were extracted from circles of a $40''$
radius centered at the pulsar position, encircling about 88\% of
all detected source counts. The extracted spectra were binned
with at least 25 source counts per bin. 
It was found that the pulsar emission is completely 
buried under background
at energies above 3 keV. Hence, only the 0.3--3 keV range
was used for the EPIC-MOS data on the pulsar (0.3 keV is the
calibration ``threshold'' for these instruments --- see \S\,2.2). 
The inferred total source count rates are $13\pm 2$ and $17\pm 2$ 
counts ks$^{-1}$ for the EPIC-MOS1 and MOS2 detectors, respectively.
The EPIC-pn source (plus background) spectrum was extracted from 
the CCD columns \#\,35--42 (see insert in Fig.~7). 
This was estimated to provide about 82\%
of all pulsar counts detected with EPIC-pn. 
Background was evaluated from the columns \#\,51--58.
The EPIC-pn spectrum of \psrb\ in the 0.3--2 keV range
was binned with at least 40 source counts in each bin.
The total source count rate in this instrument is 
$57\pm 3$ counts ks$^{-1}$.

Estimates for the hydrogen column density in the direction to
\psrb\ are rather uncertain. The pulsar dispersion measure
$DM=4.6$ pc cm$^{-3}$ pc suggests $\nh\sim 1$, whereas
estimates\footnote{
See {\tt http://archive.stsci.edu/euve/ism/ismform.html}}
for objects close to \psrb\ indicate on larger
values, $\nh\sim 3$. The total Galactic column density in the direction
to \psrb\ is $n_{\rm HI}\simeq 6\times 10^{20}$ cm$^{-2}$.
Hence, $\nh=1$--3 can be regarded as a plausible range.
The three EPIC spectra were first fitted with a single PL model.
Although this fit is formally acceptable ($\chi^2_\nu=1.1$
for $\nu=121$), it yielded a very large photon index $\Gamma=3.3\pm 0.4$
and too high hydrogen column density $\nh=16\pm 5$.
These two parameters make the pure nonthermal interpretation of 
the pulsar X-ray emission infeasible.
Applying a single thermal model, either BB or one-temperature PCs
with hydrogen atmosphere leaves a significant data excess at energies
$E\gapr 1.5$ keV, suggesting that one more
component is required to fit the spectra.

First two-component model to test was one-temperature PCs with hydrogen
atmosphere plus a PL spectrum. 
Such a model yielded a good fit with $\chi^2_\nu=1.1$ ($\nu=119$) and
hydrogen column density $\nh=3\pm 2$ compatible with
the independent estimates.
As the pulsar emission is detected
only below 3 keV, the slope of the nonthermal component is rather
unconstrained, $\Gamma=2.1\pm 0.7$. Nevertheless, the (isotropic)
nonthermal luminosity is well determined, 
$\lnon=(0.9\pm0.2)\times 10^{30}$ erg s$^{-1}$
in the 0.2--10 keV range (for $d=0.27$ kpc).
The inferred PC parameters are
$\tpc=1.3\pm 0.1$ MK and $\rpc=0.32\pm 0.04$ km (for
the standard NS mass 
$M=1.4\,M_\odot$ and radius $R=10$ km, 
and pulsar angles $\zeta=\alpha=45^\circ$).
The corresponding bolometric luminosity of two PCs,
$\lbol=(1.0\pm 0.2)\times 10^{30}$ erg s$^{-1}$,
is almost the same as the estimated PL luminosity. 
A best PC-plus-PL fit is shown in Figure~8.
As hydrogen atmosphere
spectra are much harder than the BB ones at the same effective
temperature, using the BB model for the thermal component
results in a higher temperature, $\tbb\sim 2.4$ MK, and a much smaller
radius of emitting area, $\rbb\sim 0.04$ km (see, e.g., 
Zavlin \& Pavlov 2004 for more examples).

Next, a two-temperature PC (``core''+``rim'')
model was applied to the EPIC
spectra of \psrb.  It yielded a fit of the same quality 
as the one-temperature PC plus PL fit and
the following model parameters:
$\tpcc=2.2\pm 0.2$ MK, $\tpcr=0.5\pm 0.1$ MK, and
$\rpcc=(0.11\pm 0.03)$ km, $\rpcr=(1.9\pm 0.7)$ km
(for the same assumption on the NS mass, radius and
geometry as above). 
Interestingly, these numbers are similar to those estimated for
\psra\ (\S\,2.2).
The bolometric luminosity
of these PCs is $\lbol=(1.8\pm 0.1)\times 10^{30}$ erg s$^{-1}$.
No other spectral component (e.g., PL) in addition to this 
PC model is required to fit the observed data.  
An upper limit on nonthermal emission
of \psrb\ is $\lnon<0.09\times 10^{30}$ erg s$^{-1}$ 
(in the 0.2--10 keV range). 

\subsection{Timing analysis}
To obtain a maximal signal-to-noise ratio ($S/N=21$),
the EPIC-pn CCD columns \#\,37--40 were used for extracting
source (plus background) counts in the 0.3--2 keV range.
Of total 8506 counts extracted, 39\% was estimated to
be emitted from the pulsar. 
Using the radio ephemeris parameters of \psrb\ from 
the ATNF catalog, spin frequency
$f_0=202.793897234988$ Hz and its derivative 
$\dot{f}=-8.447\times 10^{-16}$ s$^{-2}$ (at MJD~50288.0), 
in the $Z^2_n$ test immediately revealed a signal,
with the most significant
component $Z^2_2=97.1$. The probability
to obtain such a value by chance in one trial is
$4.1\times 10^{-20}$. The maximum value $Z^2_{2,{\rm max}}=97.6$
is reached at a frequency which differs 
from $f_0$ by $\delta{f}=0.7$ $\mu$Hz, 
similar to the case of \psra\ (\S\,2.3).

Because of the strong background contamination
and relatively narrow energy range where the pulsar data 
are available, only the energy-integrated (in 0.3--2 keV)
light curve of \psrb\ was extracted (see Fig.~9).
It reveals one broad prominent pulse per period
(at phase $\phi\approx 0.8$ in Fig.~9)
and a possible weaker peak separated from the main one 
by $\delta\phi\approx 0.35$ (or $\delta\phi\approx 0.65$,
depending on which pulse is leading). Alternatively,
the obtained pulse may be described as
a single broad peak with an asymmetric shape, 
a steeper rise and a longer trail, as predicted by relativistic
effects (in particular, the Doppler boost) in fast rotating
pulsars (Braje et al.\ 2000). The estimated 
source intrinsic pulsed fraction,
$\fp=56\pm 14\%$ is fairly high, but it is still consistent with
thermal PC models, depending on the NS geometry and compactness
(the $M/R$ ratio).

\section{\psrc}
This is a solitary pulsar with a spin period $P\simeq 5.2$ ms,
characteristic age $\tau\simeq 4.4$ Gyr and rotational energy
loss $\ed\simeq 5.3\times 10^{33}$ erg s$^{-1}$ (according to
the ATNF catalog). It is a relatively close object located
at a distance $d\simeq 0.39$ kpc, as derived from the pulsar
dispersion measure $DM=6.49$ cm$^{-3}$ pc
and the model Galactic distribution
of free electrons by Cordes \& Lazio (2003).
From $ROSAT$ data Becker \& Tr\"umper (1999) suggested
an X-ray counterpart for \psrc\ (at a 4$\sigma$ significance level).
Because of very poor statistics available ($\sim 25$ counts)
and properties of the instrument used in the $ROSAT$
observation, neither spectral nor timing information on
the pulsar X-ray counterpart could be obtained.

\subsection{\xmm\ observation}
\psrc\ was observed with \xmm\ on 2003 December 2  (orbit 729)
for 68.0 and 66.2 ks effective exposures for the EPIC-MOS and
EPIC-pn instruments, respectively, with the same filter and 
observational modes as for \psra. Figure~10 shows a $20'\times20'$
combined EPIC-MOS1 and MOS2 image in the 0.3-2 keV range.
\psrc\ is clearly detected at a position which differs from its
radio position by only $0\farcs4$, that is well within
the $2''$--$3''$ uncertainty of the EPIC absolute astrometry.
As seen in the EPIC-MOS image, the EPIC-pn field of view contains many
background sources, with a comparable (or even larger) brightnesses,
which heavily contaminate the 1-D image of \psrc\ in the EPIC-pn
data (see insert in Fig.~10).
This makes the EPIC-pn data 
virtually useless for spectral analysis of the pulsar emission
because no reliable background subtraction is possible.

\subsection{Spectral analysis}
Only EPIC-MOS data were used to evaluate spectral properties
of \psrc. The pulsar counts were extracted from circles
of a $20''$ radius, which contain about 70\% of all
photons detected from \psrc. Background was estimated from a few
source-free regions. The extracted spectra were binned with
at least 20 source counts per bin in the 0.3--2 keV range
and 9--11 source counts in the bins at 2--3 keV.
The total source count rates are $2.3\pm 0.3$ and $2.7\pm 0.3$ 
counts ks$^{-1}$ for the EPIC-MOS1 and MOS2 detectors, respectively.

The pulsar dispersion measure gives a hydrogen column density
$\nh\sim 2$ toward \psrc. Independent estimates of interstellar
absorption for objects close to the pulsar suggest $\nh\sim 6$,
whereas the total Galactic column density is 
$n_{\rm HI}\simeq 5\times 10^{20}$ cm$^{-2}$.
A single PL fit to the pulsar spectra produced a best result
with $\Gamma=3.7\pm 0.8$ and $\nh=23\pm 5$ ($\chi^2_\nu=1.4$ for
$\nu=11$). Similar to the case of \psrb, these
inferred parameters make the pure nonthermal interpretation
hardly plausible. A single BB model fits the spectra equally
well ($\chi^2_\nu=1.1$). The obtained parameters are $\nh=2\pm 2$
(compatible with the independent estimates),
$\tbb=2.9\pm 0.3$ MK and $\rbb=0.03\pm 0.01$ km (for $d=0.39$ kpc).
One-temperature PC model with nonmagnetic hydrogen atmosphere
(and the same assumption on the NS parameters as in \S\S\,2.2 and 3.2)
yielded $\tpc=1.8\pm 0.4$ MK and $\rpc=0.1\pm 0.1$ km
($\chi^2_\nu=1.0$),
and the same $\nh$ range as in the BB fit.
A best PC model is shown in Figure~11.
The estimated bolometric luminosity of two PCs
is $\lbol=(0.4\pm 0.2)\times 10^{30}$ erg s$^{-1}$.
Adding a PL component of $\Gamma=2$
to the best PC model fit put virtually the same upper limit on
the nonthermal flux in the 0.2--10 keV range, 
$\lnon < 0.08\times 10^{30}$ erg s$^{-1}$, 
as found for \psrb\ in the two-temperature PC interpretation
(\S\,3.2).
On the other hand, one can assume that the above-mentioned PC
model is in fact the PC ``core'' and estimate temperature and 
size of the PC ``rim'' adding second thermal component.
We found that to obtain a ``rim'' radius compatible with the
estimates found for PSRs J0437--4715 and J2124--3358,
a larger hydrogen density or smaller ``core'' radius is required.
For example, if the PC ``rim'' temperature and radius are
$\tpcr=0.4$ MK and $\rpcr=1.1$ km, then the other three parameters
can be $\nh=3$, $\tpcc=2.1$ MK and $\rpcc=0.07$ km,
or $\nh=6$, $\tpcc=1.7$ MK and $\rpcc=0.15$ km. We note that
although there exist many such combinations of model parameters,
the main contribution to the thermal luminosity, about 75\%,
comes from the softer (``core'') component as determined in the
one-temperature PC fit. 

\subsection{Timing analysis}
The pulsar radio ephemeris parameters
from the ATNF catalog, $f_0=193.71568669103$ Hz and 
$\dot{f}=-6.953\times 10^{-16}$ s$^{-2}$
(at MJD~51018.0),
were used for searching pulsations in the X-ray flux 
detected from \psrc\ with the EPIC-pn instrument.
An optimal column interval in the 1-D distribution of 
EPIC-pn counts (Fig.~10) was determined with the aid of the $Z^2_1$ test
(other harmonics with $n>1$ were found to be less significant).
It gave $Z^2_1=36.1$ for 854 source-plus-background counts
extracted from the CCD columns \#\,38--39 in the 0.3--2 keV range. 
Varying spin frequency
in vicinity of $f_0$ resulted in a maximal value 
$Z^2_{1,{\rm max}}=39.1$ at
a frequency differing by $\delta{f}=2.3$ $\mu$Hz from $f_0$.
The relative deviation, $\delta{f}/f=1.2\times 10^{-8}$,
is still consistent with the EPIC-pn timing accuracy.
This value $Z^2_{1,{\rm max}}$ corresponds to a signal
detection at a 6$\sigma$ significance level (probability
to obtain this value by chance in one trial is $3\times 10^{-9}$).

Figure~12 presents the light curve of \psrc\ extracted in the
0.3--2 keV range. The pulse profile reveals
a single broad peak per period, in agreement with the result
of the $Z^2_n$ test. To determine the intrinsic source pulsed
fraction, we first estimated how many photons of those 854 extracted
belong to the pulsar. To do this, 1-D count
distribution were constructed from EPIC-MOS photons detected 
in the EPIC-pn filed of view divided in two parts, with and without
the pulsar emission (see caption to Fig.~10), 
similar to what was done for \psra.
It gave an estimate of 34\% on the fraction of pulsar photons
in the total number of EPIC-MOS counts of $E=0.3$--2 keV
extracted in the CCD columns \#\,38--39.
Assuming that the same fraction stays for the EPIC-pn photons\footnote{
This assumption may not be very accurate because of, e.g.,
different internal background in the EPIC-MOS and EPIC-pn
instruments.}
gives the intrinsic pulsed fraction of about 52\%, 
with a rather large error.

\section{\psrd}
It is the third fastest pulsar known, with a spin period 
$P\simeq 1.9$ ms. This pulsar is in a circular binary
system with a low-mass companion and 1.6 d orbital period.
The standard estimates give its characteristic age $\tau\simeq 6$ Gyr,
rotational energy loss $\ed\simeq 3.0\times 10^{34}$ erg s$^{-1}$,
and a distance to the pulsar $d\simeq 0.54$ kpc.
In earlier years \psrd\ was observed in X-rays only with $Beppo$SAX
(in July 1999 for about 100 ks), but the pulsar was not detected.

\xmm\ was pointed at \psrd\ on 2002 June 16 (orbit 463) for 
35.0 and 32.1 ks effective exposures with the EPIC-MOS and EPIC-pn
detectors, respectively, with medium filters and the same observational
modes as in the observations of the other three pulsar described 
\S\S\,2--4.
A $20'\times 20'$ combined EPIC-MOS1 and MOS2 image is shown in
Figure~13. In this image an X-ray source is evident at a position
which differs by only $0\farcs2$ from the radio position of \psrb.
Unfortunately, in the EPIC-pn observation with the satellite position
angle of $65.5^\circ$ 
this X-ray source was completely buried under
emission from the much brighter (by a factor of 10) object indicated
with the label ``S'' in the image. Therefore, neither spectral
nor temporal information on this source could be obtained from 
the EPIC-pn
data (timing analysis of EPIC-pn counts extracted at the source
position revealed no pulsations). The count rates of the
source at the pulsar position  are 
$0.9\pm 0.5$ and $1.3\pm 0.5$ counts ks$^{-1}$
in the EPIC-MOS1 and MOS2 detectors, respectively.
Despite the low ($<3\sigma$) formal significance of the X-ray detection,
the proximity of this source to the radio position of \psrd\ makes 
it a likely X-ray counterpart of the pulsar.
About 70 counts detected from this source with the two EPIC-MOS
detectors at $E\lapr 3$ keV can be equally well
fitted with a single PL spectrum of $\Gamma\sim 2.5$ or 
a single BB model of $\tbb\sim 2.2$ MK and $\rbb\sim 0.05$ km
(for $d=0.54$ kpc).
Both these models yields an X-ray luminosity $L_X\sim
0.4\times 10^{30}$ erg s$^{-1}$.

\section{Discussion and conclusion}
The new \xmm\ observations have enabled us to perform
the spectral and timing analyses of the X-ray data
collected from three pulsars,
J0437--4715, J2124-3358, and J1024--0719. In addition,
a possible X-ray counterpart of \psrd\ has been found,
with an estimate on its X-ray flux.
For the first three objects the \xmm\ data have revealed
strong evidences of presence of thermal components in
their X-ray emission. These components are interpreted
as radiation emitted from heated PCs around magnetic
poles on the NS surface. 

In the case of \psra, the new data
confirmed the main result obtained from the previous observations
that the pulsar spectrum is best fitted with
a two-temperature PCs model with nonmagnetic hydrogen
atmosphere plus a nonthermal (PL) component which contributes
mainly at energies $E\gapr 3$ keV.
In this interpretation, the PC temperature changes from about
$\sim 1.5$ MK in the PC ``core'' of a $\sim 0.4$ km radius
down to $\sim 0.5$ MK in the PC ``rim'' of a $\sim 2.5$ km radius.
The \xmm\ data allowed us to investigate the 
energy-resolved pulse profiles in a broad range of $E=0.3$--6 keV,
as well as perform phase-resolved spectroscopy. This analysis
showed that the pulse shape is different at lower and higher
energies (below and above $\sim 2$ keV), with pulsed fraction $\fp$
increasing by about 10\% in the 0.3--2 keV range where the
thermal emission dominates. This increase of $\fp$
can be explained by the anisotropic properties of radiation emitted
from a NS hydrogen atmosphere (see Zavlin et al.\ 1996, and
Zavlin \& Pavlov 1998 for more details). The pulsed fraction
in the light curve extracted at energies above 2 keV is rather
large, $\fp\approx 60\%$. In addition, the peak in the
pulse profile obtained at higher energies is narrower than
the pulse in the low-energy light curve and shifted by about
0.1 in phase. These properties provide strong evidence that two different
emission mechanisms generate the X-ray emission of \psra.

The \xmm\ spectral data on \psrb\ can be most plausibly 
interpreted as either a combination of a one-temperature PC model
plus a nonthermal PL spectrum or a pure thermal radiation
from PCs with a nonuniform temperature decreasing from $\sim 2$ MK
to $\sim 0.5$ MK in the ``core'' and ``rim'' areas of the PCs,
similar to that derived for \psra. 
Comparing the PC radii inferred in these two approaches ---
about 0.3 km with the PC-plus-PL combination and 1--2 km in the 
two-temperature PC model --- with
the canonical estimate of pulsar models,
$\rpc^*=R(2\pi R/cP)^{1/2}\simeq 2.1$ km 
for \psrb\
(assuming the NS radius $R=10$ km), suggests that the pure thermal
interpretation is preferable. In any case, the 
thermal luminosities derived in these model fits differ insignificantly,
by only a factor of $\sim 2$. In addition, the new data allowed us
to firmly detect pulsations of the X-ray flux of \psrb\
at the radio pulsar period, although the shape of the obtained 
pulse profile cannot be interpreted unambiguously. 
It may be described as a profile with two peaks of different
heights separated by about 0.35 (or 0.65) in phase. If so,
it would suggest that two different components are detected in
the pulsar emission. On the other hand, it could be a broad
single peak produced by thermal PC emission with a
shape distorted by the relativistic effects
in fast rotating NSs.

For \psrb, the \xmm\ observation provided the first 
firm detection of the pulsar X-ray emission. Although
the obtained statistics is much poorer than that delivered
in the observations of \psrb\ and, especially \psra,
it has shown that the pulsar's spectrum is best described as a 
pure thermal radiation emitted from PCs with either uniform
or nonuniform (``core''+``rim'') temperature.
The derived total thermal luminosity is almost 
independent on assumption about the PC temperature
distribution. In addition to the spectral analysis, the new data 
revealed pulsations of the pulsar X-ray flux at a highly
significant level. Most likely, the pulsar light curves shows a 
single broad
pulse per rotational period, although much better statistics
is required to draw more definitive conclusions on the pulse shape.

Regarding the \xmm\ observation of \psrd, only a possible
X-ray counterpart of the pulsar could be suggested based
on the very close proximity of this source to the pulsar
position. A longer observation, with a proper
satellite position angle, is needed to perform a meaningful
investigation of X-ray radiation of this pulsar.

With a sample of millisecond pulsars showing
thermal PC emission available (see Table~2), it is natural
to compare observational properties of the PC radiation, first of all,
the luminosities, with those predicted by pulsar models.
A most recent and detailed analysis of PC heating is presented
in Harding \& Muslimov (2001, 2002) who modified 
the space charge limited flow
model by Arons (1981) to account for the relativistic effects 
and electric field screening above the NS surface.
In these models the PC heating is produced by 
inverse Compton scattering (ICS)
of thermal X-rays from the NS surface by primary electrons
accelerated in the pulsar magnetosphere, as well as by
positrons generated through curvature radiation (CR) and
returning to the NS surface from the upper pair formation front.
These authors showed that the PC heating 
from the CR pair fronts dominates for most pulsars except
those with very short spin periods and low magnetic fields
(see Fig.~1 in Harding \& Muslimov 2002) because the latter
objects cannot produced CR pairs.
Therefore, the ICS is the only mechanism to heat PCs
of millisecond pulsars. 
Harding \& Muslimov (2002) calculated the ``PC efficiency'' of one PC,
$\epc=\lbolpc/\ed$, for typical parameters of millisecond
pulsars (age $\tau=0.1$--10 Gyr, spin period $P=2$--5 ms).
This model predicts $\epc$ ranging between about 
$5\times 10^{-7}$ and $5\times 10^{-5}$ for PCs with temperatures 
$\tpc=1$--3 MK (see Fig.~8 in Harding \& Muslimov 2002). 
Comparing these estimates with those derived
from the observations (Table~2) shows that the maximum predicted value
of $\epc_{\rm max}$ is lower by a factor of 10 than the 
results obtained for
PSRs J0437--4715 and J0030+0451, and a factor of 6 for
\psrb. This difference is larger
for PSR~J1012+5307, but we note that the $\epc$ value
given in Table~2 for this pulsar was obtained under assumption
that all X-ray flux detected from PSR~J1012+5307 is of a thermal
origin. For \psrc, the value of $\epc_{\rm max}$
well agrees with the observational result.
On the other hand, the predicted PC luminosities, 
$\lbolpc\sim 10^{29}$--$10^{30}$ erg s$^{-1}$ for $\tpc=1$ MK
and up to $10^{31}$ erg s$^{-1}$ for $\tpc=3$ MK
(see Fig.~9 in Harding \& Muslimov 2002),
are in much better agreement with those estimated for the objects
listed in Table~2. To conclude, the PC heating from ICS pairs
is a promising model for interpreting properties of thermal
PC emission observed from millisecond pulsars.
 
Finally, as mentioned in \S\,1, there is no clear understanding yet
at which conditions the thermal PC component is expected to dominate over 
the nonthermal one in X-ray radiation of millisecond pulsars. 
Saito et al.\ (1997) first suggested that high luminosity
of the nonthermal emission may be associated with 
large values of the magnetic field at the pulsar light cylinder,
$B_{\rm lc}=B_{\rm surf} [R/R_{\rm lc}]^3$ 
($R_{\rm lc}=cP/2\pi$). Indeed, the four pulsars
emitting pure nonthermal emission, PSRs B1937+21, B1957+20,
B1821--24, and J0218+4232 (see \S\,1 and Table~1),
possess magnetic fields,
$B_{\rm lc}\sim (0.3$--$1)\times 10^6$ G, close to that
of the Crab pulsar ($\simeq 1\times 10^6$ G) and
exceeding those in the millisecond pulsars 
with thermal emission, $B_{\rm lc}\sim (2$--$3)\times 10^4$ G,
at least by an order of magnitude. This provides a strong indication
that the nonthermal emission of millisecond pulsars is generated in
emission zone(s) close to the light cylinder (as predicted
by the outer-gap pulsar models --- see, e.g., Cheng et al.\ 1986)
and $B_{\rm lc}$ is one of the main parameters to govern the
magnetospheric activity.

\acknowledgements
The author thanks George Pavlov for helpful and stimulating
discussions.
This work is supported by a National Research Council
Research Associateship Award at NASA MSFC.


\clearpage

\begin{deluxetable}{lllllll}
\tabletypesize{\scriptsize}
\tablewidth{0pt}
\tablecaption{Properties of nine millisecond pulsars\tablenotemark{a}}
\tablehead{\colhead{PSR} & \colhead{$P$} & \colhead{$d$} & 
\colhead{$\tau$} & \colhead{$\log\ed$} & \colhead{$\log L_X$} & Refs. \\
 & \colhead{(ms)} &  \colhead{(kpc)} & \colhead{(Gyr)}  & 
\colhead{(erg\,s$^{-1}$)} & \colhead{(erg\,s$^{-1}$)} & }
\startdata
B1937+21       & 1.56 & 3.57 &  0.24 & \,\,36.04 & \,\,33.15 & 1,2 \\
B1957+20       & 1.61 & 2.49 &  2.24 & \,\,35.20 & \,\,31.81 & \,\,3 \\
J0218+4232     & 2.32 & 2.67 &  0.48 & \,\,35.38 & \,\,32.54 & \,\,4 \\
B1821--24      & 3.05 & 3.09 &  0.03 & \,\,36.34 & \,\,32.71 & \,\,5 \\
J0751+1807     & 3.48 & 1.15 &  7.08 & \,\,33.86 & \,\,30.84 & \,\,6 \\
J0030+0451     & 4.87 & 0.32 &  7.71 & \,\,33.53 & \,\,30.40 & \,\,7 \\
J2124--3358    & 4.93 & 0.27 &  3.80 & \,\,33.83 & \,\,30.23 & \,\,8 \\
J1012+5307     & 5.26 & 0.41 &  4.86 & \,\,33.67 & \,\,30.38 & \,\,6 \\
J0437--4715    & 5.76 & 0.14 &  4.89 & \,\,33.58 & \,\,30.46 & \,\,9 \\
\enddata
\tablenotetext{a}{
Based on results obtained in previous works
(see references below).
}
\tablecomments{
Second through fourth columns give standard pulsar parameters:
spin periods $P$, distances $d$, characteristic ages $\tau$,
and rotation energy losses $\ed$. 
All distances are estimated from the pulsar dispersion measures
and the model of Galactic distribution of free electrons
(Cordes \& Lazio 2003), except for 
\psra\ with the distance determined from
the pulsar's parallax (van Straten et al.\ 2001).
The fifth column provides pulsars' X-ray luminosities in 
the 0.2--10 keV range as derived in the corresponding references 
(sixth column):
1 -- Takahashi et al. (2001),
2 -- Nicastro et al. (2004),
3 -- Stappers et al. (2003),
4 -- Webb et al. (2004a),
5 -- Becker et al. (2004),
6 -- Webb et al. (2004b),
7 -- Becker \& Aschenbach (2002),
8 -- Sakurai et al. (2001)
9 -- Zavlin et al. (2002)
}
\end{deluxetable}
\clearpage


\begin{deluxetable}{llll}
\tabletypesize{\scriptsize}
\tablewidth{0pt}
\tablecaption{Thermal luminosities of millisecond pulsars}
\tablehead{\colhead{PSR} & \colhead{$\log\ed$} & \colhead{$\log\lbolpc$} 
& \colhead{$\log\epc$} \\
 & \colhead{(erg\,s$^{-1}$)} & \colhead{(erg\,s$^{-1}$)} & }
\startdata
J0437--4715    &  \,\,33.58 & \,\,30.23 & \,-3.35 \\
J2124--3358    &  \,\,33.83 & \,\,29.96 & \,-3.87 \\
J1024--0719    &  \,\,33.72 & \,\,29.27 & \,-4.45 \\
%
%
J1012+5307\tablenotemark{a}     &  \,\,33.67 & \,\,30.52 & \,-3.15  \\
J0030+0451     &  \,\,33.53 & \,\,30.18 & \,-3.35 \\
\enddata
\tablecomments{
The thermal (bolometric) luminosities of one PC 
are determined as $\lbolpc=\lbol/2$, with $\lbol$ estimates
obtained in this work for the first 
three
pulsars.
For PSRs J0030+0451 and J1012+5307 the PC luminosities are
calculated as $\lbolpc=g_{\rm r}^{-2}\lbol^\infty/2$, where
$\lbol^\infty$ are luminosities derived from BB spectral fits
by Becker \& Aschenbach (2002) and Webb et al.\ (2004b),
respectively, and $g_{\rm r}=[1-2GM/c^2R]^{-1/2}$ ($=0.77$ 
for $M=1.4\,M_\odot$ and $R=10$ km).
The ``PC efficiency'' is defined as $\epc=\lbolpc/\ed$.
}
\tablenotetext{a}{
Assuming that all flux detected for this pulsar 
is of a thermal origin.
}
\end{deluxetable}
\clearpage
\newpage


\clearpage

\begin{figure}
\plotone{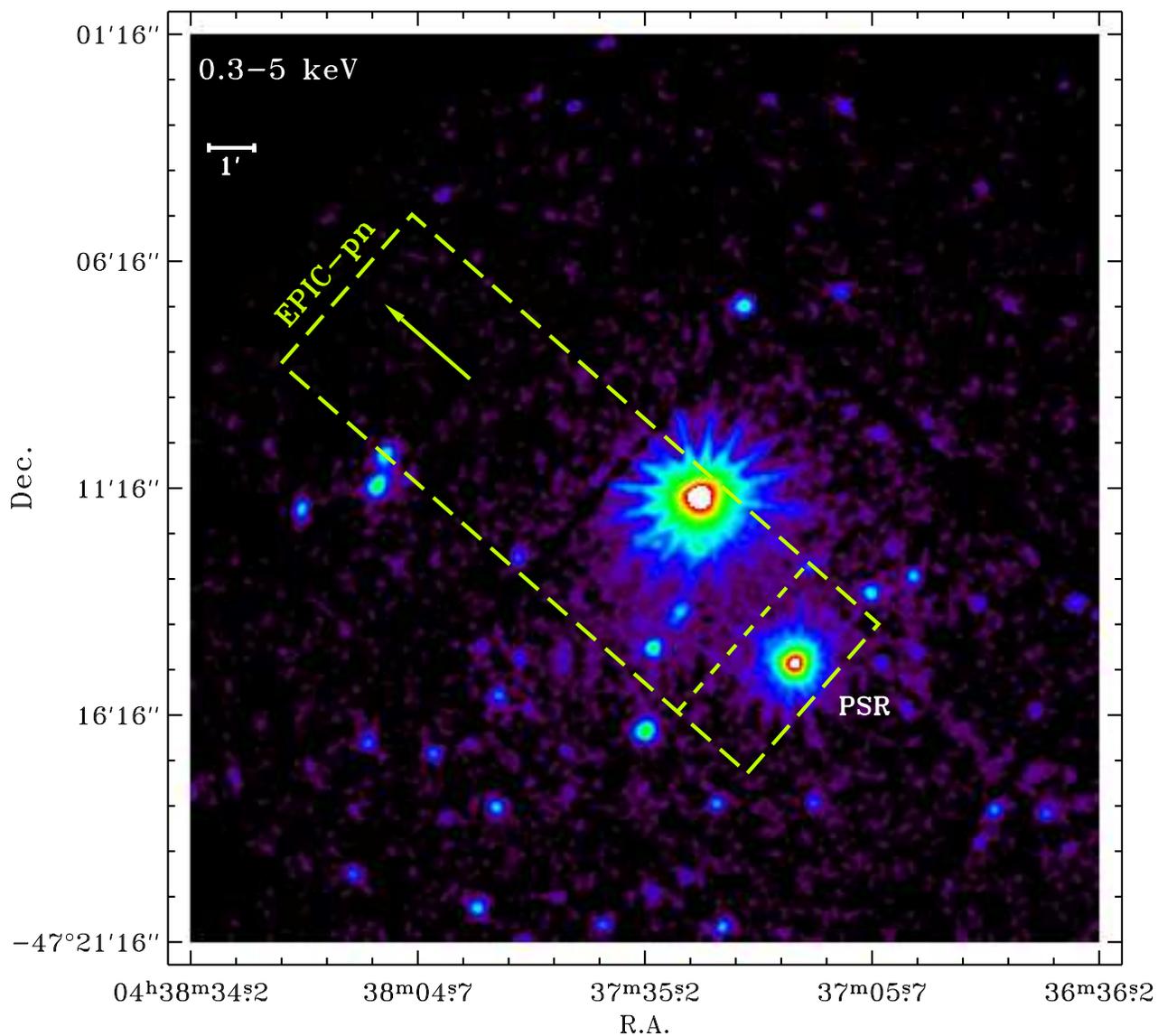}
\vskip -1cm
\caption{Combined EPIC-MOS1 and MOS2 image in the 0.3--5 keV range,
with \psra\ marked with the label ``PSR''. Long-dashed rectangle
depicts the field of view of the EPIC-pn CCD used in the 
\xmm\ observation of the pulsar.
Arrow shows the direction of the CCD read-out
perpendicular to which EPIC-pn counts were collapsed
in a 1-D  distribution (see Fig.~2). 
Short-dashed line separates the areas 
within the EPIC-pn field of view with negligible contributions
from the pulsar and AGN emission
(above and below the line, respectively).
}
\end{figure}
\clearpage

\begin{figure}
\plotone{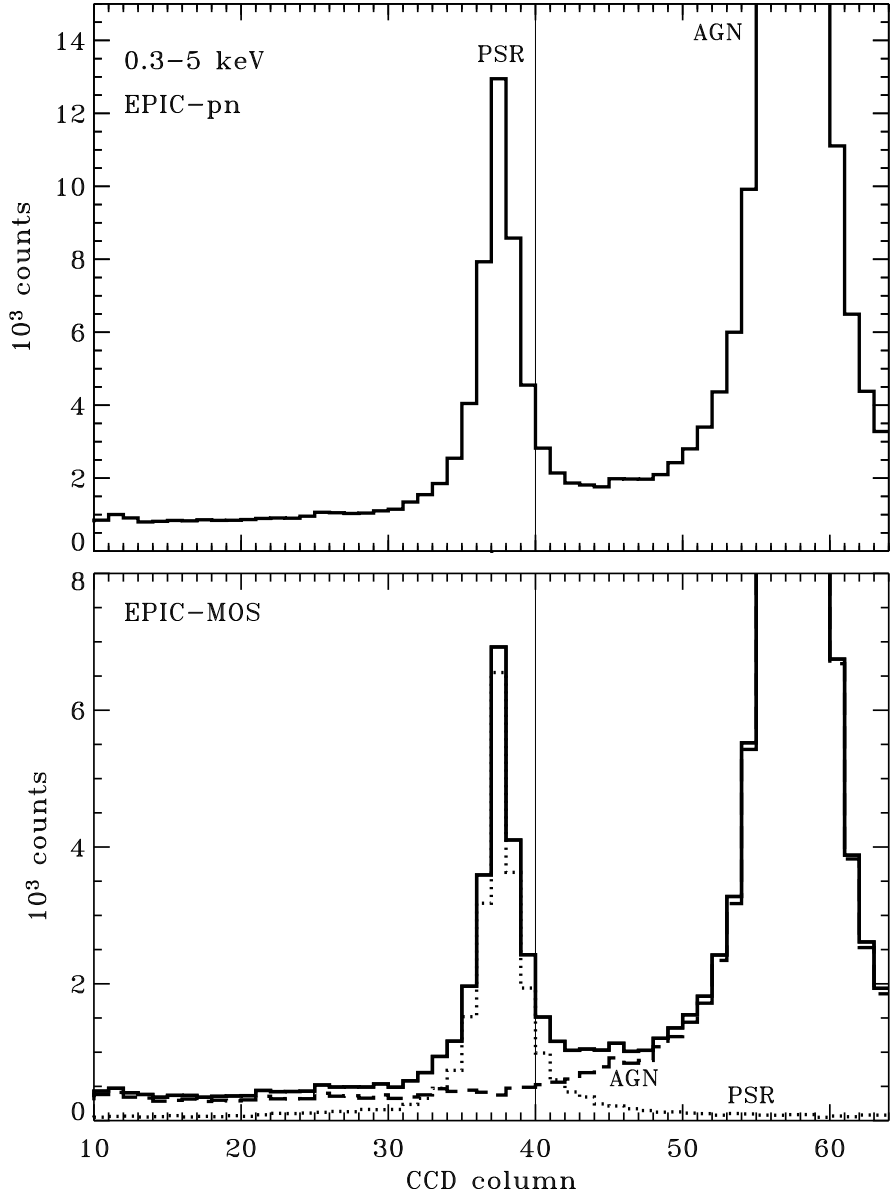}
\vskip -1cm
\caption{{\it Upper panel:} 1-D distribution
of EPIC-pn counts in the 0.3-5 keV range.
Peaks labeled with ``PSR'' and ``AGN'' are the images of \psra\
and the neighboring AGN (see Fig.~1 and text). One CCD column 
has a width of $4\farcs2$.
{\it Lower panel:} 1-D image (solid line) 
constructed from the combined EPIC-MOS1 and
MOS2 data confined within the EPIC-pn filed of view 
drawn in Fig.~1. Dotted (labeled with ``PSR'') and dashed 
(labeled with ``AGN'') histograms show 1-D EPIC-MOS images computed for
the lower and upper areas of the EPIC-pn field of view (below and
above the short-dashed line in Fig.~1, respectively). 
Thin vertical lines in the both
panels indicate the CCD column below which contamination of the pulsar
emission by that from the AGN is negligible.
}
\end{figure}
\clearpage

\begin{figure}
\plotone{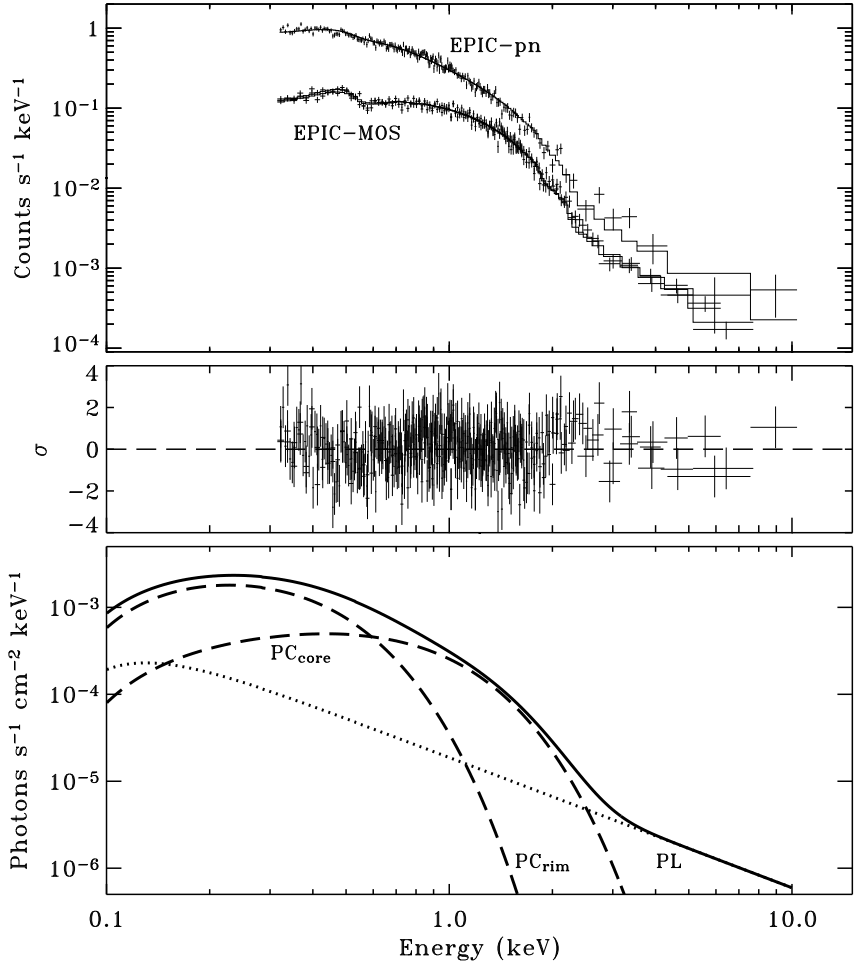}
\vskip -1cm
\caption{EPIC-pn and two EPIC-MOS phase-integrated
spectra of \psra\ fitted with
a best two-temperature polar cap (PC) model (``core''+``rim'')
plus a power-low (PL) component, 
residuals of the fit, and model fluxes
({\it upper, middle}, and {\it lower panels,} respectively).
}
\end{figure}
\clearpage

\begin{figure}
\plotone{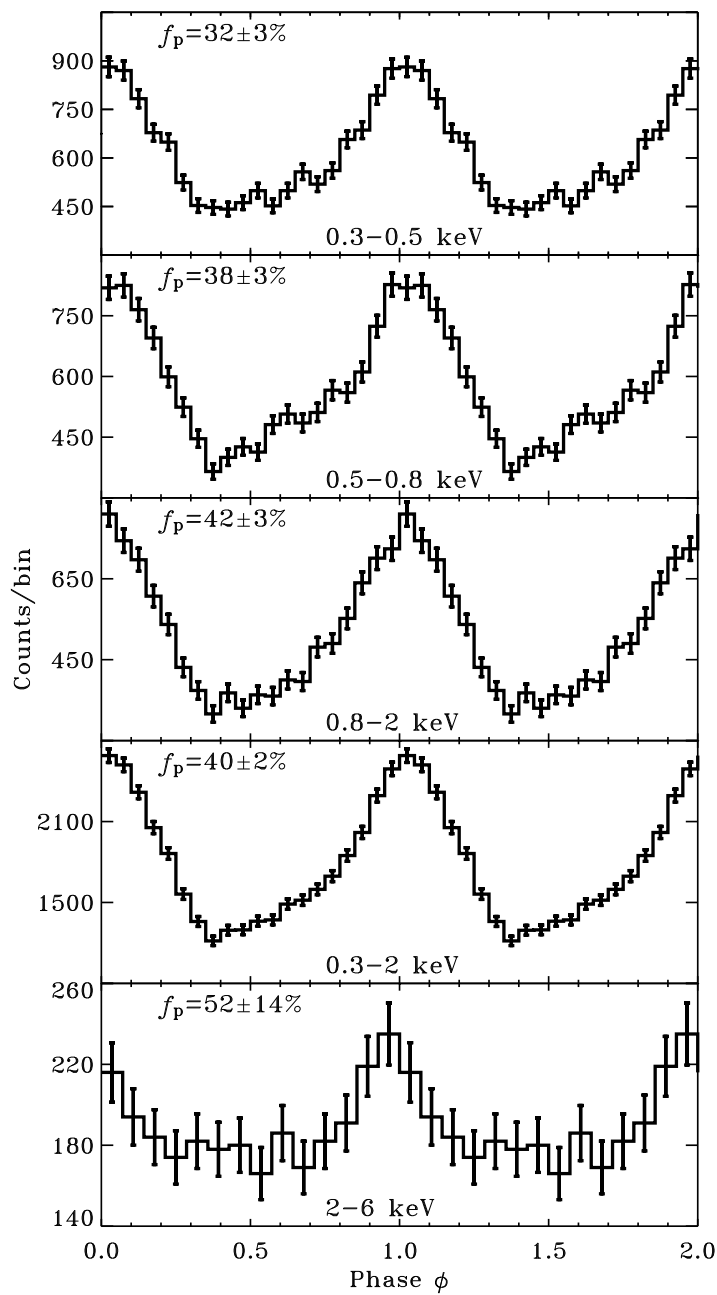}
\vskip -1cm
\caption{Energy-resolved pulsed profiles of \psra\
extracted from the EPIC-pn data, with estimated values
of the intrinsic source pulsed fraction $\fp$.
}
\end{figure}
\clearpage

\begin{figure}
\plotone{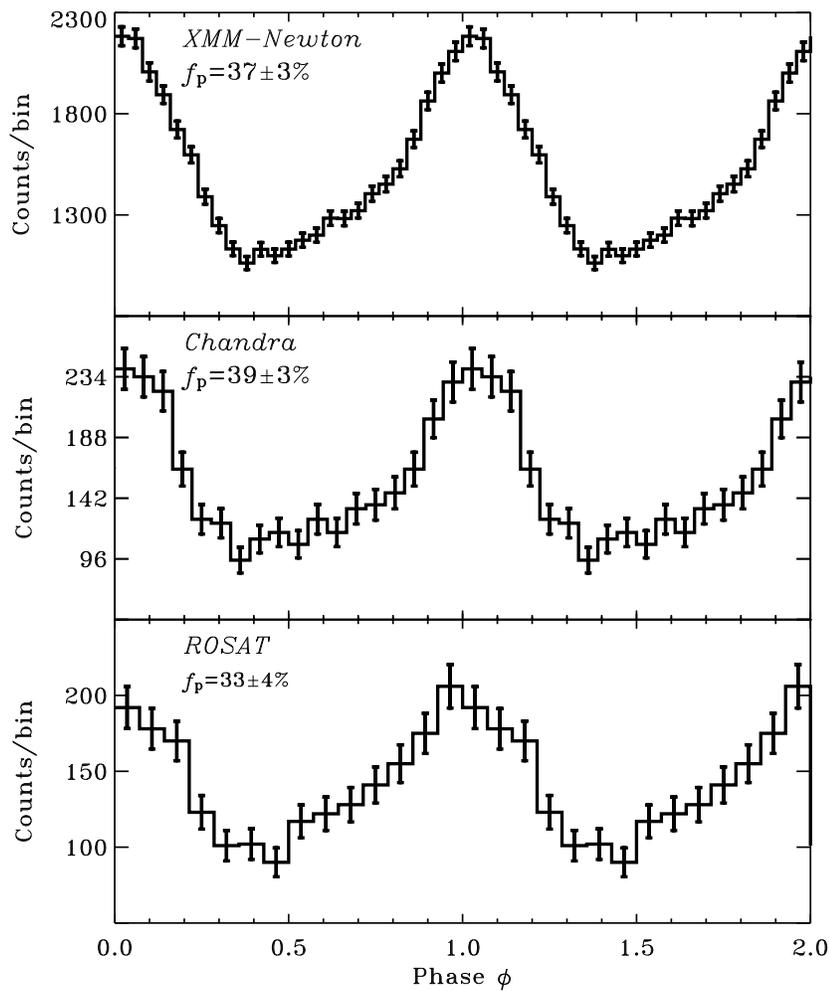}
\vskip -1cm
\caption{Pulsed profiles of \psra\
obtained from the \xmm\ (0.3--6 keV), $Chandra$ 
(0.1--10 keV) and $ROSAT$ (0.1--2.4 keV) data,
with estimated values of the intrinsic source pulsed fraction $\fp$.
Zero phases are arbitrary.
}
\end{figure}
\clearpage

\begin{figure}
\plotone{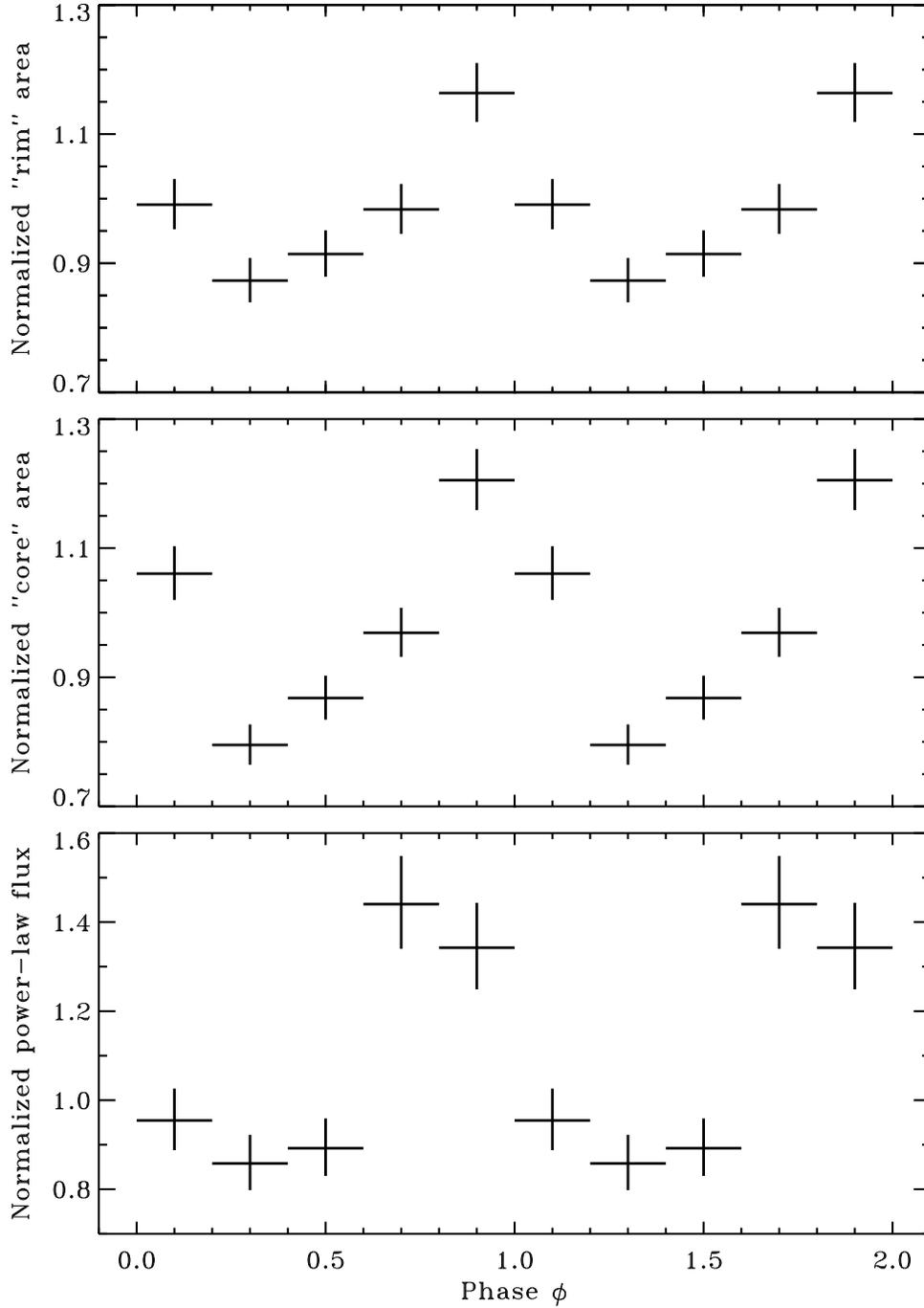}
\vskip -1cm
\caption{Phase dependences of the polar cap (PC) areas
and nonthermal (PL) flux in the two-temperature
PC (``core''+``rim'') plus PL model.
}
\end{figure}
\clearpage

\begin{figure}
\plotone{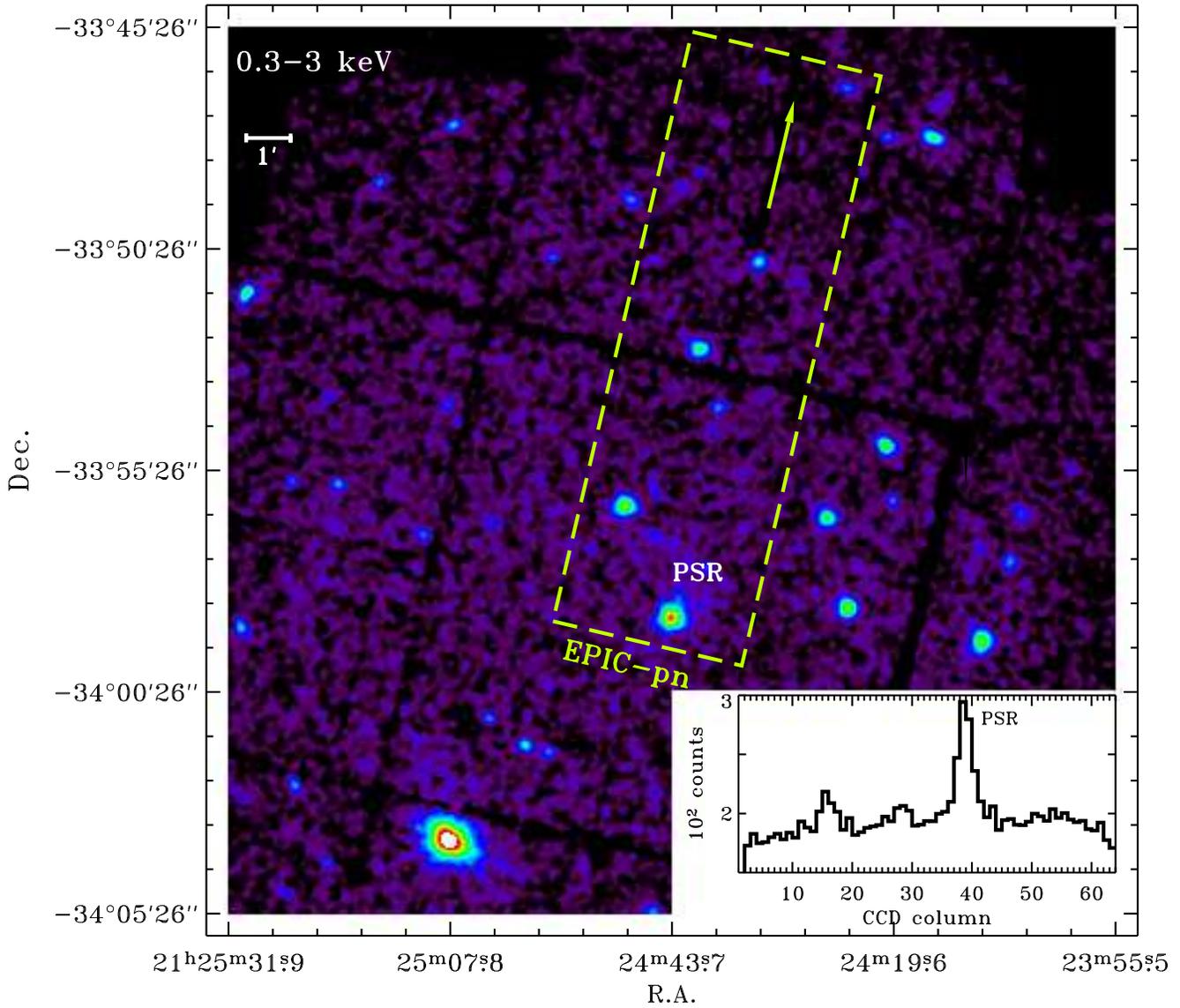}
\vskip -1cm
\caption{
Combined EPIC-MOS1 and MOS2 image in the 0.3--3 keV range,
with \psrb\ marked with the label ``PSR'' 
(see caption to Fig.~1 for other details).
One-dimensional distribution of EPIC-pn counts is shown in the
insert.
}
\end{figure}
\clearpage

\begin{figure}
\plotone{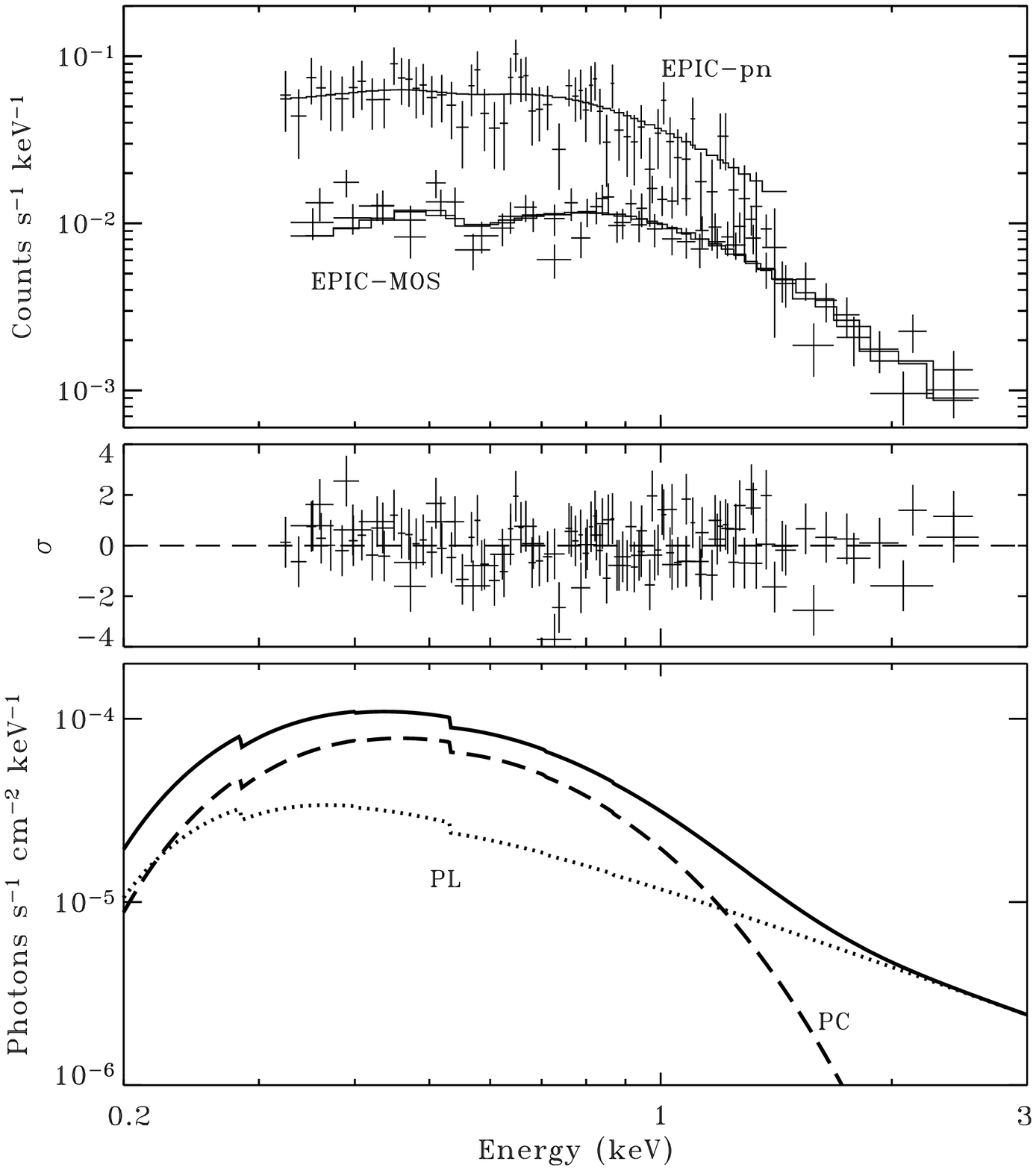}
\vskip -1cm
\caption{
EPIC-pn and two EPIC-MOS phase-integrated spectra of \psrb\ fitted with
a two-component (polar cap [PC]
plus a power-low [PL]) model,
residuals of the fit, and model fluxes
({\it upper, middle,} and {\it lower panels,} respectively).
}
\end{figure}
\clearpage

\begin{figure}
\plotone{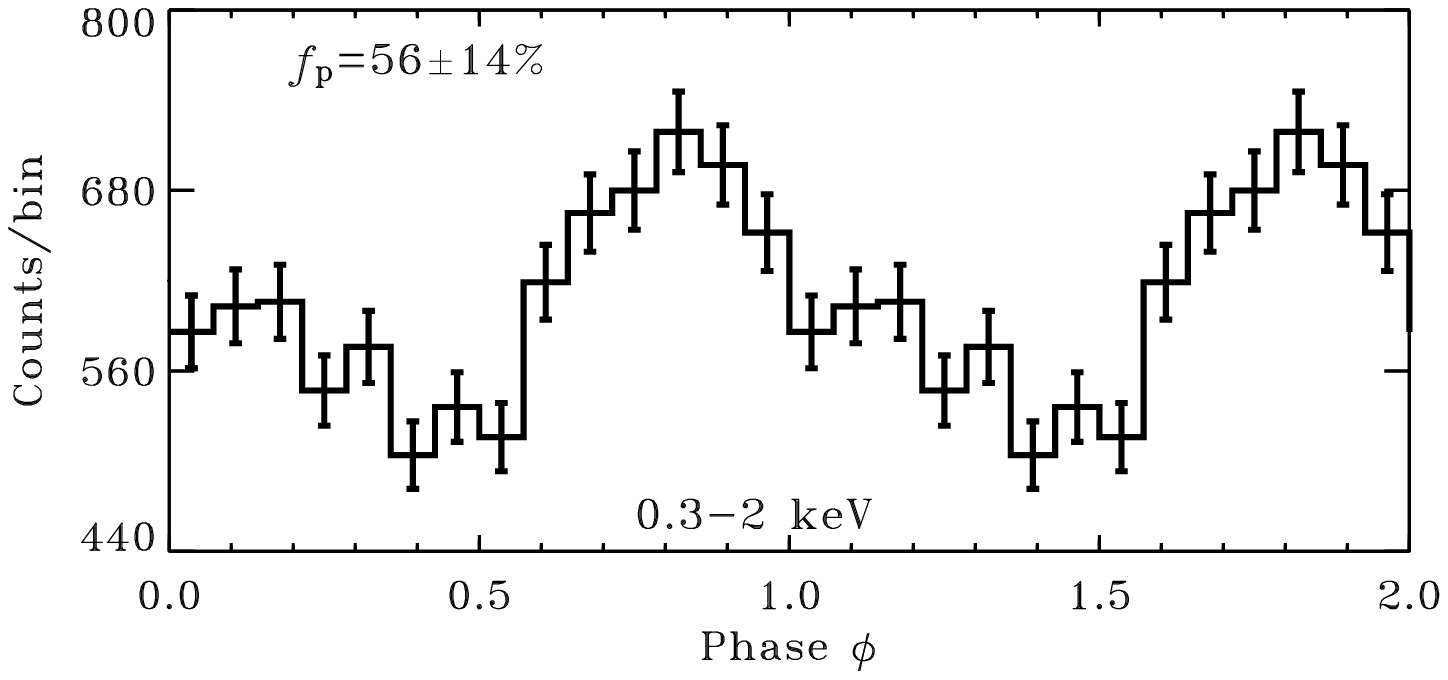}
\vskip -1cm
\caption{Pulse profile of \psrb\
extracted from the EPIC-pn data in the 0.3--2 keV range, 
with estimated value of the intrinsic source pulsed fraction $\fp$.
}
\end{figure}
\clearpage

\begin{figure}
\plotone{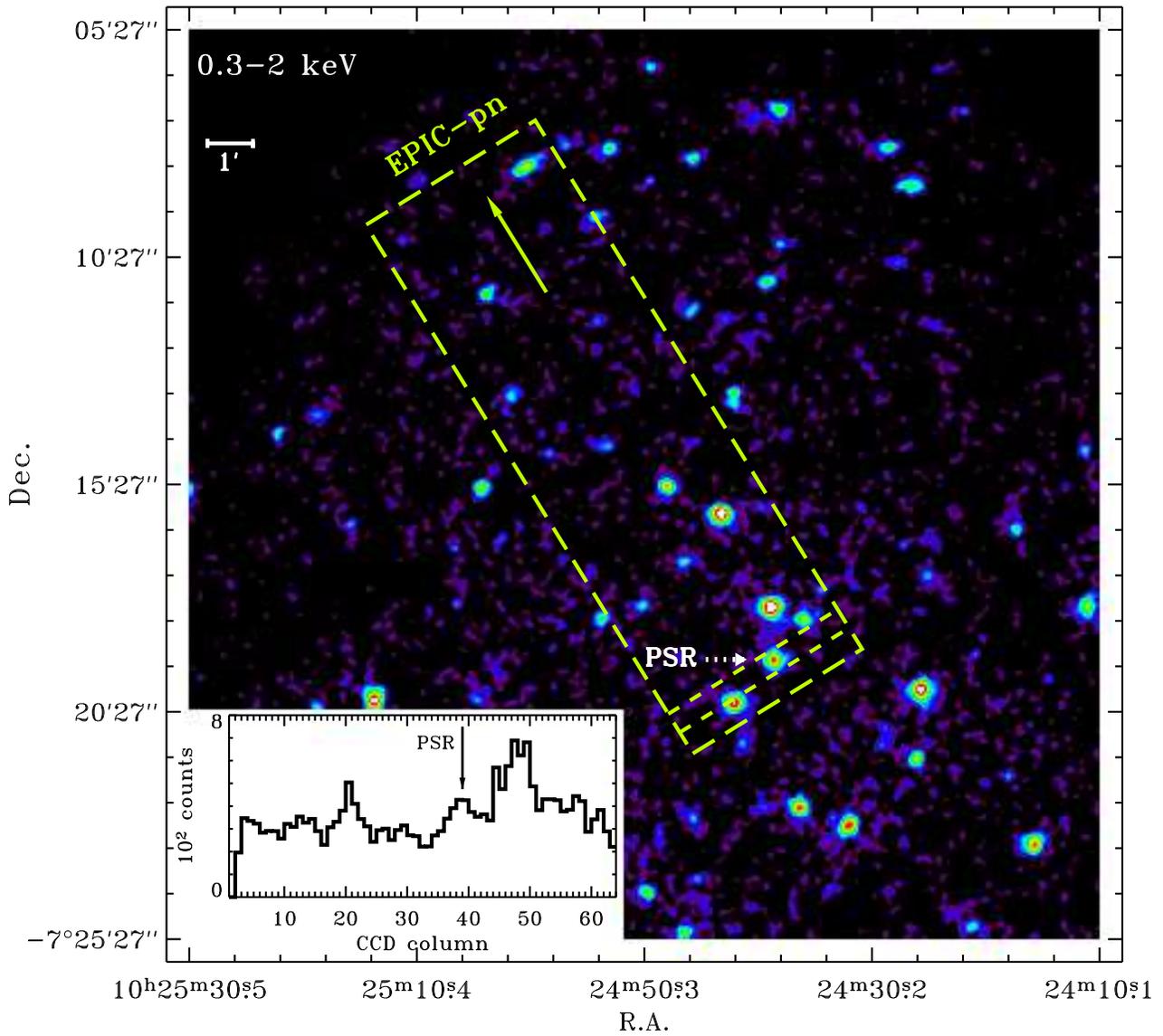}
\vskip -1cm
\caption{
Combined EPIC-MOS1 and MOS2 image in the 0.3--5 keV range,
with \psrc\ indicated with dotted arrow
(see caption to Fig.~1 for other details).
Emission of the pulsar dominates in the area between 
two short-dashed lines within the EPIC-pn field of view.
One-dimensional distribution of EPIC-pn counts is shown in the
insert. 
}
\end{figure}
\clearpage

\begin{figure}
\plotone{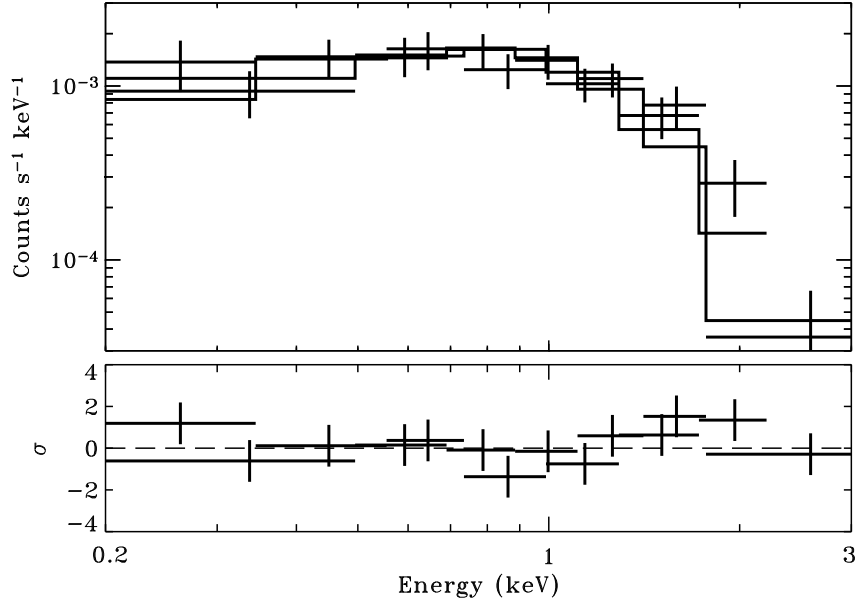}
\vskip -1cm
\caption{
Two EPIC-MOS phase-integrated spectra of \psrc\ fitted with
a polar cap model, and
residuals of the fit.
}
\end{figure}
\clearpage

\begin{figure}
\plotone{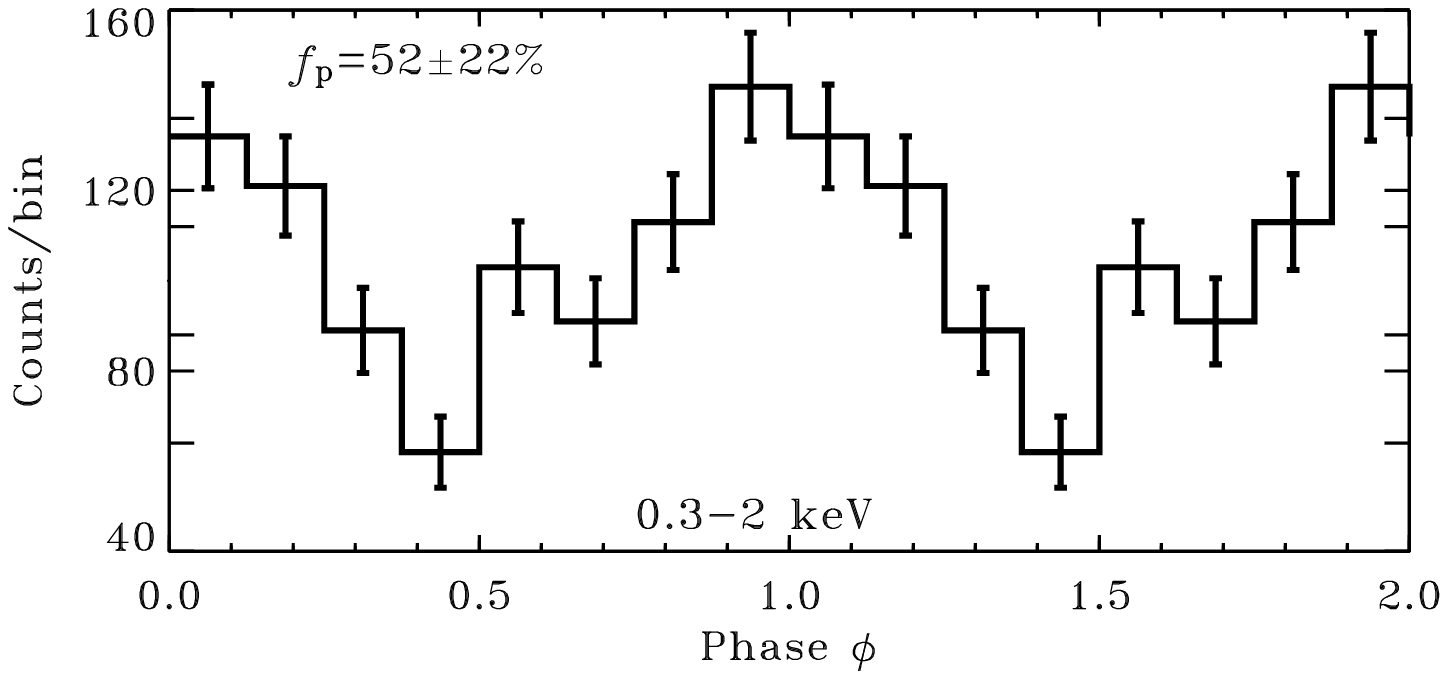}
\vskip -1cm
\caption{Pulse profile of \psrc\
extracted from the EPIC-pn data in the 0.3--2 keV range,
with estimated value of the intrinsic source pulsed fraction $\fp$.
}
\end{figure}
\clearpage

\begin{figure}
\plotone{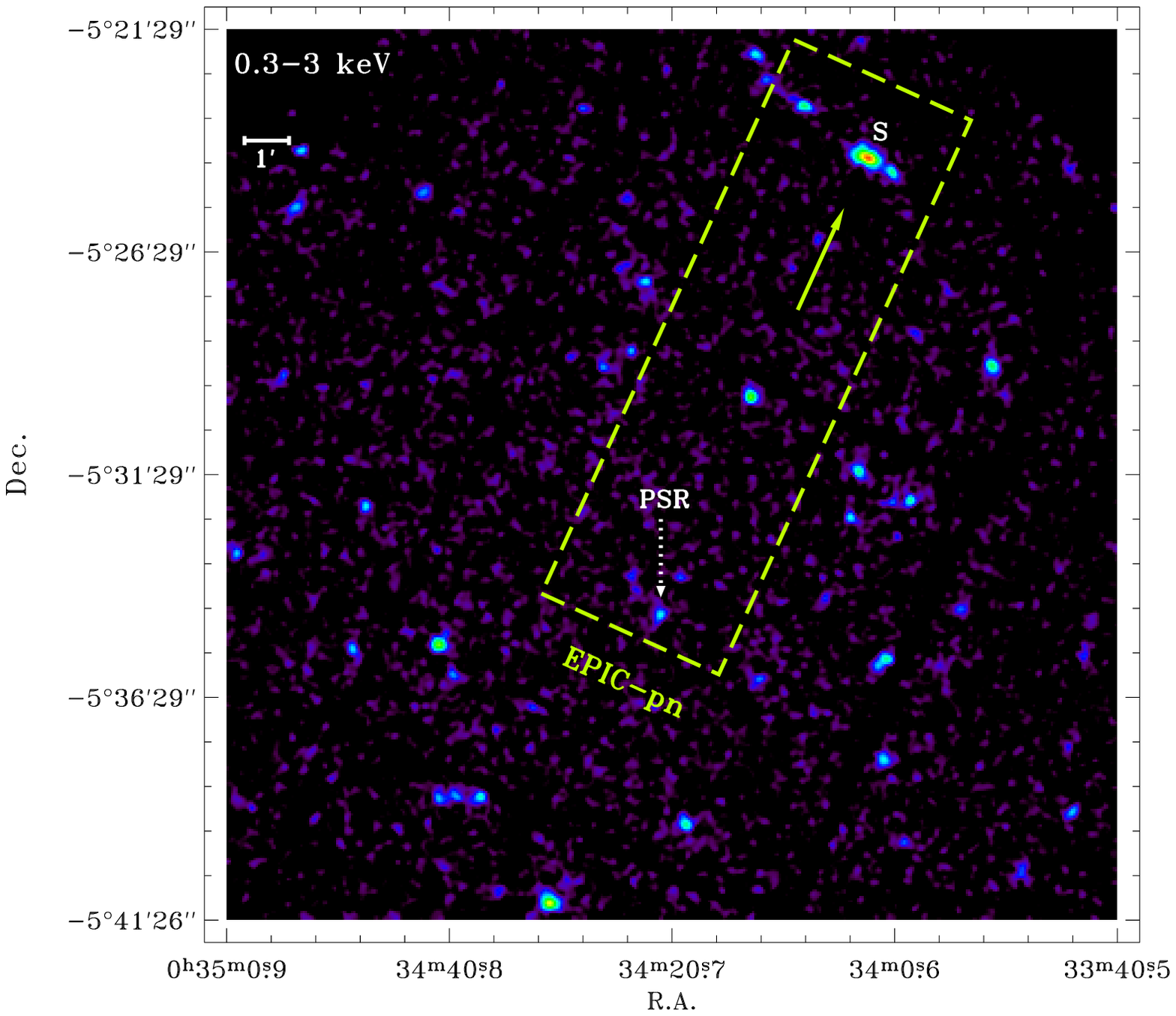}
\vskip -1cm
\caption{
Combined EPIC-MOS1 and MOS2 image in the 0.3--3 keV range,
with \psrd\ indicated with dotted arrow
(see caption to Fig.~1 for other details).
The brightest source in the filed of view is marked with
the label ``S''.
}
\end{figure}
\clearpage



\begin{thebibliography}{}

\bibitem[]{1180} Arons, J. 1981, ApJ, 248, 1099

\bibitem[]{998} Becker, W., \& Tr\"umper 1993, Nature, 365, 528

\bibitem[]{1000} Becker, W., \& Tr\"umper 1999, A\&A, 341, 803

\bibitem[]{1002} Becker, W., \& Aschenbach, B. 2002,
in Neutron Stars, Pulsars and Supernova Remnants,
Proc. of the 270-th Heraeus Seminar, ed. W. Becker, H. Lesch \&
J. Tr\"umper (MPE Report 278), 64

\bibitem[]{1007} Becker, W., et al. 2003,
ApJ, 594, 798

\bibitem[]{1185} Buccheri, R., et al. 1983, A\&A, 128, 245

\bibitem[]{1012} Braje, T.\ M., Romani, R.\ W., \& Rauch, K.\ P.
 2000, ApJ, 531, 447

\bibitem[]{1015} Cheng, K.\ S., Ho, C.\, Ruderman, M. 1986, ApJ, 300, 500

\bibitem[]{1017} Cordes, J.\ M., \& Lazio, T.\ J.\ W. 2003,
preprint (astro-ph/0301598).

\bibitem[]{1020} Grindlay, J.\ E., Camilo, F., Heinke, C.\ O.,
Edmonds, P.\ D., Cohn, H., \& Lugger, P. 2002, ApJ, 581, 470

\bibitem[]{1199} Harding, A.\ K., \& Muslimov, A.\ G. 2001, ApJ, 556, 987 

\bibitem[]{1201} Harding, A.\ K., \& Muslimov, A.\ G. 2002, ApJ, 568, 862 

\bibitem[]{1027} Kargaltsev, O., Pavlov, G.\ G., \&
Romani, R.\ W. 2004, ApJ, 602, 327

\bibitem[]{} Kargaltsev, O.\ Y., Pavlov, G.\ G., Zavlin, V.\ E., \&
Romani, R.\ W. 2005, ApJ, 625, 307

\bibitem[]{1030} Kaspi, V.\ M., Roberts, M.\ S.\ E., \& Harding, A.\ K.
2006, in Compact Stellar X-ray Sources, ed.
W.\ H.\ G. Lewin \& M. van der Klis, in press (astro-ph/0402135)

\bibitem[]{1034} Kuiper, L., \& Hermsen, W. 2004,
in X-ray and Gamma-ray Astrophysics of Galactic Sources,
Proc. of the 4-th AGILE Science Workshop,
ed. M Tavani, A. Pellizzoni \& S. Vercellone, p.~47
(astro-ph/0312204)

\bibitem[]{1038} Nicastro, L., Cusumano, G., L\"ohmer, O.,
Kramer, M., Kuiper, L., Hermsen, W., Mineo, T., \&
Becker, W. 2004, A\&A, 413, 1065

\bibitem[]{1042} Pavlov, G.\ G., Zavlin, V.\ E., \& Sanwal, D. 2002,
in Neutron Stars, Pulsars and Supernova Remnants,
Proc. of the 270-th Heraeus Seminar, ed. W. Becker, H. Lesch \&
J. Tr\"umper (MPE Report 278), 283

\bibitem[]{1047} Saito, Y., Kawai, N., Kamae, T., Shibata, S.,
Dotani, T., \&Kulkarni, S.\ N. 1997, ApJL, 477, 37

\bibitem[]{1050} Sakurai, I., Kawai, N., Torii, K., Negoro, H.,
Nagase, F., Shibata, S., \& Becker, W. 2001, PASJ, 53, 535

\bibitem[]{1053} Stappers, B.\ W., Gaensler, B.\ M.,
Kaspi, V.\ M., van der Klis, M., \& Lewin, W.\ H.\ G. 2003,
Science, 299, 1372

\bibitem[]{1057} Takahashi, M., et al.
2001, ApJ, 554, 316

\bibitem[]{1060} Taylor, J.\ H., Manchester, R.\ N., \&
Lyne, A.\ G. 1993, ApJS, 88, 529

\bibitem[]{1063} Webb, N.\ A., Olive, J.-F., \& Barret, D. 2004a,
A\&A, 417, 181

\bibitem[]{1066} Webb, N.\ A., Olive, J.-F., Barret, D., Kramer, M.,
Cognard, I., \& L\"ohmer, O. 2004b, A\&A, 419, 269

\bibitem[]{1069} Zavlin, V.\ E., \& Pavlov, G.\ G. 1998, 329, 583

\bibitem[]{1071} Zavlin, V.\ E., \& Pavlov, G.\ G. 2004, ApJ, 616, 452

\bibitem[]{1073} Zavlin, V.\ E., \& Pavlov, G.\ G., \& Shibanov,
Yu.\ A. 1996, A\&A, 315, 141

\bibitem[]{1076} Zavlin, V.\ E., Pavlov, G.\ G., Sanwal, D.,
Manchester, R.\ N., Tr\"umper, J., Halpern, J.\ P., \& Becker, W.
2002, ApJ, 569, 894


\end{thebibliography}
\end{document}